\documentclass[aps,prd,12point,twocolumn,nofootinbib,showpacs,superscriptaddress]{revtex4-1}

\usepackage{graphicx}
\usepackage{dcolumn}
\usepackage{tabu}
\usepackage{amsmath,amssymb}
\usepackage{float}
\usepackage{mathrsfs}
\usepackage{accents}
\usepackage{bm}
\usepackage{lipsum}
\usepackage[dvipsnames]{xcolor}
\usepackage[title]{appendix}
\usepackage{soul}
\usepackage[breaklinks=true,pdfborder={0 0 0},bookmarksopen]{hyperref}
\hypersetup{pdfstartview={XYZ null null 1.00}}
\hypersetup{
	colorlinks   = true, 
	urlcolor     = blue, 
	linkcolor    = blue, 
	citecolor   = red 
}

\begin{document}

\title{Revisiting wormhole energy conditions in Riemann-Cartan spacetimes and under Weyl transformations}

\author{Fay\c{c}al Hammad} \email{fhammad@ubishops.ca} 
\affiliation{Department of Physics and Astronomy \& STAR Research Cluster, Bishop's University, 2600 College Street, Sherbrooke, QC, J1M~1Z7 Canada} 
\affiliation{Physics Department, Champlain 
College-Lennoxville, 2580 College Street, Sherbrooke,  
QC, J1M~0C8 Canada}

\author{\'Etienne Mass\'e}
\email[]{Etienne.Masse@usherbrooke.ca}
\affiliation{Department of Physics and Astronomy \& STAR Research Cluster, Bishop's University, 2600 College Street, Sherbrooke, QC, J1M~1Z7
Canada} 
\affiliation{D\'epartement de Physique, Universit\'e de Sherbrooke, Sherbrooke, QC, J1K~2X9 Canada}

\author{Patrick Labelle}
\email[]{plabelle@crc-lennox.qc.ca}
\affiliation{Physics Department, Champlain 
College-Lennoxville, 2580 College Street, Sherbrooke,  
QC, J1M~0C8 Canada}

\begin{abstract}
We show that the violation of the null energy condition by matter, required by traversable wormholes, can be removed in spacetimes with torsion. In addition, we show that this violation can also be removed in the conformal frame obtained by a Weyl transformation. This comes about because both conformally transformed wormholes and wormholes in spacetimes with torsion become sustained thanks to a combination of ``normal'' matter, that satisfies the null energy condition, and an induced ``geometric'' energy-momentum tensor that is able to greatly violate the null energy condition.
\end{abstract}

\maketitle
\section{Introduction}\label{sec:1}
Wormholes arose as extremely important theoretical possibilities within general relativity \cite{ER}. However, the discovery that, to be sustained, wormholes require the violation of the null energy condition (NEC), and hence the presence of the so-called ``exotic'' matter \cite{MorrisThorne,Visser,Lobo,HochbergVisser}, pushed these objects further into the realm of pure theoretical curiosities. Just like for the possibility of curing the Big Bang singularity \cite{Hehl,Kopczy'nski1972,Kopczy'nski1973,Trautman,Tafel1973,Tafel1975,Kuchowicz1975A,Kuchowicz1975B,Kuchowicz1975C,Kuchowicz1976A,Kuchowicz1976B,Kuchowicz1978,Gasperini,Poplawski}, however, one might hope that by adding torsion to spacetime, i.e., by using Riemann-Cartan spacetimes, the need for exotic matter to make traversable wormholes might also be removed.

In fact, various authors have investigated the issue and came to the following different but complementary conclusions: ({\it i}\,) torsion might provide the required degree of ``exoticity" \cite{Anchordoqui} ({\it ii}\,) a non-minimal coupling between torsion and matter might relieve the latter from violating the NEC \cite{JawadRani} ({\it iii}\,) for specific values of the matter spin traversable wormholes might definitely be produced without having matter violate the NEC \cite{Battista}. These possibilities can actually be understood from general backgrounds as follows. 

Recall that a given form of matter is said to satisfy the NEC if its energy-momentum tensor $T_{ab}^M$ satisfies the inequality $T_{ab}^M\xi^a\xi^b\geq0$ for any null vector $\xi^a$. Thus, thanks to Einstein field equations $G_{ab}\equiv R_{ab}-\tfrac{1}{2}g_{ab}R=T^M_{ab}$\footnote{Throughout the paper, we set $8\pi G=\hbar=c=1$.}, the NEC is also equivalent to the following statement in terms of the Ricci tensor: $R_{ab}\xi^a\xi^b\geq0$. It is specifically the violation of this latter inequality that is required by traversable wormholes and that consequently leads to the violation of the NEC by matter. 
In the presence of torsion, however, additional degrees of freedom are added to spacetime and the field equations governing the latter are different from Einstein equations as they contain on the right-hand side extra terms arising from torsion. In fact, the field equations, called Einstein-Cartan-Sciama-Kibble field equations, may be given a form similar to those of general relativity:
$R_{ab}-\tfrac{1}{2}g_{ab}R=T_{ab}^M+({\text{torsion terms}})_{ab}$. (See e.g., Ref. \cite{SabbataGasperini} for a textbook introduction or the very comprehensive review in Ref.~\cite{HehlReview}). It is then clear that the requirement $R_{ab}\xi^a\xi^b<0$, imposed by traversable wormholes, would lead instead to an inequality of the form $[T_{ab}^M+({\text{torsion terms}})_{ab}]\xi^a\xi^b<0$. The presence of $({\text{torsion terms}})_{ab}$ besides the energy-momentum tensor $T_{ab}^M$ inside the square brackets means that matter does not necessarily have to violate the NEC for the full inequality to hold.

Now, a very similar reasoning actually applies when investigating the fate of the NEC around the throat of a traversable wormhole built from a given traversable one by subjecting the host spacetime to a Weyl transformation. In fact, under a Weyl transformation Einstein equations also acquire an extra term and take the form, $\tilde{G}_{ab}=\tilde{R}_{ab}-\tfrac{1}{2}\tilde{g}_{ab}\tilde{R}=T_{ab}^M+T^{\rm Induced}_{ab}$ \cite{DabrowskiAl}. The induced energy-momentum tensor comes from the deformation of the spacetime metric caused by the Weyl transformation. Thus, a traversable wormhole, requiring $\tilde{R}_{ab}\tilde{\xi}^a\tilde{\xi}^b<0$ in the conformal frame, would require an inequality of the form $(T_{ab}^M+T^{\rm Induced}_{ab})\tilde{\xi}^a\tilde{\xi}^b<0$. Thanks to this extra term, we see indeed that matter does not also necessarily have to violate the NEC for the full inequality to hold.

In Ref.~\cite{HochbergVisser} it has been proven on general grounds how traversable wormholes do require the violation of the NEC. In addition, it was shown in that reference that torsion does not remove the violation of the NEC but aggravates it instead. Our aim in this paper is to revisit such a derivation for the case of spacetimes with torsion, and then under a Weyl transformation, to show that the above hand-waving arguments, that clash with the conclusions of Ref.~\cite{HochbergVisser} but seem to agree with those of Refs.~\cite{Anchordoqui,JawadRani,Battista}, could actually be implemented more rigorously. In so doing, we are going to pinpoint a key fact in Riemann-Cartan spacetimes that has been missed in Ref.~\cite{HochbergVisser} and that led the authors to such an incomplete conclusion about the NEC for wormholes in Riemann-Cartan spacetimes.
Nevertheless, the rigorous study provided here shows that the conclusions are more subtle than what the above hand-waving arguments seem to suggest and thus cannot be as simple as those obtained in Ref.~\cite{Battista} either.

The paper is organized as follows. In Sec.~\ref{sec:2}, we expose the various tools and equations necessary for the description of a spacetime wormhole and the derivation of the NEC requirement. We also include briefly in the second part of that section the derivation given in Ref.~\cite{HochbergVisser} for the violation of the NEC in spacetimes with torsion and point out the subtle issue in such a derivation. In the third part of that section we derive the correct Raychaudhuri equation for null vectors in Riemann-Cartan spacetimes and use it to rigorously examine the fate of the NEC in such spacetimes. In Sec.~\ref{sec:3}, we investigate the fate of the violation of the NEC by traversable wormholes in the conformal frame, both in Riemann and in Riemann-Cartan spacetimes. We end this paper with a brief conclusion section to summarize and discuss our findings.     
\section{In the original frame}\label{sec:2}
\subsection{Without torsion}
We devote this first subsection to the main definitions used in {\it torsion-free} spacetimes for the description of traversable wormholes and the proof that traversable wormholes require the violation of the null energy condition as given in more detail in Ref.~\cite{HochbergVisser}.

A wormhole is characterized by its throat which, in turn, is best described through the behavior of light rays on that throat and in its vicinity as those rays propagate in the background spacetime hosting the wormhole. In fact, to make the throat of a traversable wormhole the spacetime should focus ingoing light rays toward the minimal surface, representing the throat, from both sides of the latter and defocus them away from the minimal surface from both sides as well. 

Mathematically, this is expressed \cite{HochbergVisser} using the expansion parameter $\theta_{\pm}$ of the geodesic congruence of the ingoing and outgoing light rays. Given the symmetry of the description, however, we are going to consider only either one, and therefore choose to use the ``neutral" expansion $\theta$ which would stand for both. In a spacetime with metric $g_{ab}$, one can always define a transverse metric $h_{ab}$ with respect to a given null direction $\xi^a$ defined as the tangent to a congruence of null geodesics. Such a transverse metric is given by\footnote{For a nice textbook introduction to the notation used and the formulas presented in this subsection, refer to Refs.~\cite{Wald,Poisson}.},
\begin{equation}\label{TransverseMetric}
h_{ab}=g_{ab}+\xi_aN_b+\xi_bN_a,
\end{equation}
where, $N_a$ is the auxiliary null vector of which the normalization is chosen such that $\xi_aN^a=-1$. It is easy to check that $h_{ab}h^{ab}=2$ and $\xi^ah_{ab}=0=N^ah_{ab}$. By taking the trace of the projection ${\mathbf B}_{ab}\equiv {h^c}_a{h^d}_bB_{cd}$ of the ``deviation" tensor $B_{ab}\equiv\nabla_b\xi_a$ on the transverse space given by this transverse metric, one obtains the expansion parameter,
\begin{equation}\label{Expansion}
\theta=g^{ab}{\mathbf B}_{ab}.
\end{equation}
By extracting the anti-symmetric part and then the symmetric-traceless part of the same projection of $B_{ab}$ on the transverse metric, one obtains, respectively, the twist tensor $\omega_{ab}$ and the shear tensor $\sigma_{ab}$ as follows,
\begin{align}
\omega_{ab}&={\mathbf B}_{[ab]},\label{Twist}\\
\sigma_{ab}&={\mathbf B}_{(ab)}-\tfrac{1}{2}\theta h_{ab},\label{Shear}\\
{\mathbf B}_{ab}&=\sigma_{ab}+\omega_{ab}+\tfrac{1}{2}\theta h_{ab}.\label{SpatialB}
\end{align}
The evolution of the expansion $\theta$ in terms of an affine parameter $\lambda$ along the null direction $\xi^a$ of the congruence is governed by Raychaudhuri's equation, which reads \cite{Wald,Poisson},
\begin{equation}\label{TorsionFreeRaychau}
\tfrac{\rm d}{{\rm d}\lambda}\theta=-\tfrac{1}{2}\theta^2-\sigma_{ab}\sigma^{ab}+\omega_{ab}\omega^{ab}-R_{ab}\xi^a\xi^b.
\end{equation}
Then, using Einstein equations $R_{ab}-\frac{1}{2}g_{ab}R=T^M_{ab}$, and taking into account that $\xi^a$ is a null vector, one can express Raychaudhuri's equation in terms of the energy-momentum tensor of matter $T^M_{ab}$ responsible for creating such a wormhole as follows, 
\begin{equation}\label{TorsionFreeRaychauBis}
\tfrac{\rm d}{{\rm d}\lambda}\theta+\tfrac{1}{2}\theta^2+\sigma_{ab}\sigma^{ab}=\omega_{ab}\omega^{ab}-T^M_{ab}\xi^a\xi^b.
\end{equation}

The fact that the throat of a traversable wormhole is, by definition, supported by a minimal surface in the spacetime is expressed mathematically by the two conditions $\theta=0$ and ${\rm d}\theta/{\rm d}\lambda>0$, both to be satisfied on the throat \cite{HochbergVisser}. On the other hand, given that a null vector is automatically hypersurface orthogonal, we conclude that the twist tensor $\omega_{ab}$ vanishes. This is known as Frobenius' theorem (see e.g., Ref.~\cite{Poisson} for a detailed derivation). In addition, the product $\sigma_{ab}\sigma^{ab}$ is everywhere positive given the spacelike nature of the shear tensor as it follows from Eq.~(\ref{Shear}). Taking into account these three facts in Eq.~(\ref{TorsionFreeRaychauBis}), the latter implies that,
\begin{equation}\label{NECProof}
T^M_{ab}\xi^a\xi^b=-\tfrac{\rm d}{{\rm d}\lambda}\theta-\sigma_{ab}\sigma^{ab}<0.
\end{equation}
This is what proves the necessity of the violation of the NEC for making a traversable wormhole \cite{HochbergVisser}.
\subsection{With torsion}
In this subsection we expose the steps followed in Ref.~\cite{HochbergVisser} to argue for the aggravation of the violation of the NEC by wormholes in spacetimes {\it with torsion}. Riemann-Cartan spacetimes are characterized by a non-symmetric connection $C_{ab}^c$, the symmetric part of which is different even from the Christoffel connection $\Gamma_{ab}^c$, in contrast to what is stated in Ref.~\cite{HochbergVisser}. This connection provides one with the covariant derivative and the torsion (notice that we are using here the convention of Ref.~\cite{DeyAl} for torsion),
\begin{align}
&\nabla_a\xi^b=\partial_a\xi^b+C_{ac}^b\xi^c,\label{TorsionCovariantDerivative1}\\
&\nabla_a\xi_b=\partial_a\xi_b-C_{ab}^c\xi_c,\label{TorsionCovariantDerivative2}\\
&C_{ab}^c-C_{ba}^c\equiv{T^c}_{ab}.\label{TorsionConnection}
\end{align}
The antisymmetric part ${T^c}_{ab}$ of the connection is a tensor, antisymmetric in its last two indices, and is called the torsion tensor. Note also, that throughout the paper the metricity condition $\nabla_ag_{bc}=0$ will be assumed.  

Raychaudhuri's equation in the presence of torsion, as given in Ref.~\cite{HochbergVisser}, was, unfortunately, based again on the same deviation tensor $B_{ab}=\nabla_b\xi_a$ of torsion-free spacetimes. The equation thus found in Ref.~\cite{HochbergVisser} has the form,
\begin{align}\label{TorsionRaychau1}
\tfrac{\rm d}{{\rm d}\lambda}\theta=&-\tfrac{1}{2}\theta^2-\sigma_{ab}\sigma^{ab}+\omega_{ab}\omega^{ab}-\accentset{\circ}{R}_{ab}\xi^a\xi^b\nonumber\\
&-2T^{abc}B_{ca}\xi_b+{T^{ab}}_cT_{abd}\xi^c\xi^d.
\end{align}
Here, $\accentset{\circ}{R}_{ab}$ is the Riemannian (torsion-free) Ricci tensor. 
At this point, the authors introduced a simplifying assumption to get to the main conclusion faster. The simplifying assumption was to choose ${T^c}_{ab}$ to be a totally antisymmetric tensor. This consists, for example, in identifying it with the Kalb-Ramond field of string theory built from the potential two-form $A_{ab}$, i.e., $H=dA$. Because of such a total antisymmetry of the tensor, the authors substituted $B_{ca}$ in Eq.~(\ref{TorsionRaychau1}) by the twist tensor $\omega_{ca}$. The authors then argued that, because $\xi^a$ is a null vector, i.e., a hypersurface orthogonal vector, one has $\omega_{ab}=0$ thanks to Frobenius's theorem \cite{Poisson}. This allowed them to remove both terms containing $\omega_{ab}$ from Eq.~(\ref{TorsionRaychau1}). Finally, the authors used Einstein-Cartan-Sciama-Kibble metric field equations in the form $\accentset{\circ}{G}_{ab}=T^M_{ab}+3T_{acd}{T_b}^{cd}-\tfrac{1}{2}g_{ab}T_{cde}T^{cde}$, where the left-hand side is the torsion-free Einstein tensor, thanks to which they turned Eq.~(\ref{TorsionRaychau1}) into \cite{HochbergVisser},
\begin{equation}\label{TorsionRaychau2}
\tfrac{\rm d}{{\rm d}\lambda}\theta+\sigma_{ab}\sigma^{ab}=-T^M_{ab}\xi^a\xi^b-2{T^{ab}}_cT_{abd}\xi^c\xi^d.
\end{equation}
The total antisymmetry of torsion allows one to express it in the form, $T_{cab}=\epsilon_{cabd}v^d$, with $\epsilon_{abcd}$ being the totally antisymmetric Levi-Civita tensor and $v^d$ being an arbitrary vector \cite{HochbergVisser}. With this form, the last term in Eq.~(\ref{TorsionRaychau2}) reads $2{T^{ab}}_cT_{abd}\xi^c\xi^d=2(v^a\xi_a)^2$. Thus, given that the left-hand side of Eq.~(\ref{TorsionRaychau2}) is positive, one finds $T^M_{ab}\xi^a\xi^b<-2(v^a\xi_a)^2$, from which it was concluded in Ref.~\cite{HochbergVisser} that torsion enhances the degree of violation of the NEC. Moreover, the authors have argued that this should hold even without the assumption of total antisymmetry for torsion.

In the next subsection, we shall see that the issue with this derivation is threefold. The first, resides in using the same deviation tensor $B_{ab}$ as in torsion-free spacetimes. The second, is ignoring the fact that torsion allows hypersurface orthogonal vectors to acquire a non-zero twist $\omega_{ab}$ \cite{DeyAl}. The third issue is the form of the above Einstein-Cartan-Sciama-Kibble field equations used in the derivation. By dealing with these issues, the conclusion one arrives at becomes completely reversed with respect to that given in Ref.~\cite{HochbergVisser} in the sense that the violation of the NEC by matter might become completely removed.
\subsection{Revisiting the original frame with torsion}
Our first aim in this subsection is to show that in Riemann-Cartan spacetimes the twist tensor does not indeed vanish even for hypersurface orthogonal vectors. Let us therefore reexamine Frobenius' theorem that proves the vanishing of $\omega_{ab}$ in torsion-free spacetimes by adapting the steps followed in its derivation \cite{Poisson} to the case of spacetimes with torsion. 

In the derivation of Frobenius' theorem, one starts by showing that the product $\xi_{[a}\nabla_b\xi_{c]}$ vanishes identically for hypersurface orthogonal vectors $\xi^a$. In the presence of torsion, however, it is easy to see that such a product does not vanish. The proof of this claim is already nicely presented in the Appendix B of Ref.~\cite{DeyAl}. The result, after putting back a missing factor of $\tfrac{1}{6}$ in the expression given in Ref.~\cite{DeyAl}, is,
\begin{equation}\label{Frobenius1}
\xi_{[a}\nabla_b\xi_{c]}=-\tfrac{1}{6}\left(\xi_a{T^d}_{bc}+\xi_c{T^d}_{ab}+\xi_b{T^d}_{ca}\right)\xi_d.
\end{equation}
Unlike Ref.~\cite{DeyAl}, in which this result has been used to indirectly infer that in general $\omega_{ab}\neq0$, we are going here to arrive at the same conclusion by finding instead an explicit expression for $\omega_{ab}$ in terms of torsion. This explicit expression will indeed serve us later in this section when dealing with Raychaudhuri's equation in relation to wormholes.

Start by taking into account that in Riemann-Cartan spacetimes the deviation tensor $B_{ab}$ is not simply equal to $\nabla_b\xi_a$, but is given instead by \cite{DeyAl},
\begin{equation}\label{BTensor}
B_{ab}=\nabla_b\xi_a+\xi^cT_{acb}.
\end{equation}
From this, we can easily deduce, after using the fact that $\xi^a$ is everywhere a null vector, and hence $\xi^a\nabla_a\xi^b=0$ for an affine parametrization, and $\xi^a\nabla_b\xi_a=0$, that,
\begin{equation}\label{XiB}
\xi^aB_{ac}=\xi^a\xi^bT_{abc}\quad \text{and}\quad\xi^aB_{ca}=0.
\end{equation}
On the other hand, the definitions (\ref{TransverseMetric}) and (\ref{BTensor}), together with identities (\ref{XiB}), provide us at once with the relation between the purely spatial tensor ${\mathbf B}_{ab}$ and the full tensor $B_{ab}$, from which we also deduce the relation between $\omega_{ab}$ and $B_{ab}$, as well as the product ${\mathbf B}_{ab}{\mathbf B}^{ba}$: 
\begin{align}
{\mathbf B}_{ab}&=B_{ab}+N^c\left(\xi_bB_{ac}+\xi_aB_{cb}\right)+N_a\xi^e\xi^cT_{ecb}\label{SpatialBFullB}\nonumber\\
&\;\;\;+N^d\left(\xi_a\xi_bN^cB_{cd}+N_a\xi_b\xi^e\xi^cT_{ecd}\right),\\
\omega_{ab}&=B_{[ab]}+N^c\left(\xi_bB_{[ac]}+\xi_aB_{[cb]}\right)\nonumber\\
&\;\;\;+\xi^e\xi^c\left(T_{ec[b}N_{a]}+N^dT_{ecd}N_{[a}\xi_{b]}\right)\label{omegaB},\\
{\mathbf B}_{ab}{\mathbf B}^{ba}&=B_{ab}B^{ba}\nonumber\\
&\;\;\;+\xi_a\xi_bN^d\left(2B_{cd}T^{abc}+N_cT^{abc}\xi^e\xi^fT_{efd}\right).\label{SpatialBSquared}
\end{align}

Next, besides expression (\ref{Frobenius1}), we can also find, thanks to the definition (\ref{BTensor}) and after recalling that ${T^c}_{ab}=-{T^c}_{ba}$, an alternative expression for $\xi_{[a}\nabla_b\xi_{c]}$ in terms of both $B_{ab}$ and ${T^c}_{ab}$ that reads, 
\begin{align}
\xi_{[a}\nabla_b\xi_{c]}=&\;\tfrac{1}{3}\left(\xi_a\nabla_{[b}\xi_{c]}+\xi_c\nabla_{[a}\xi_{b]}+\xi_b\nabla_{[c}\xi_{a]}\right)\nonumber\\=&\;\tfrac{1}{3}\left(\xi_aB_{[cb]}+\xi_cB_{[ba]}+\xi_bB_{[ac]}\right)\nonumber\\
&+\tfrac{1}{3}\xi^d\left(\xi_aT_{[cb]d}+\xi_cT_{[ba]d}+\xi_bT_{[ac]d}\right).\label{IntermediateFrobenius2}
\end{align}
By contracting both sides of identities (\ref{Frobenius1}) and (\ref{IntermediateFrobenius2}) by $N^a$ and then comparing with Eq.~(\ref{omegaB}) we easily obtain the sought-after expression for $\omega_{ab}$ purely in terms of torsion as follows,
\begin{align}\label{FinalTwist}
\omega_{ab}=&\;\xi_c\left(\tfrac{1}{2}{T^c}_{ab}-{T_{[ab]}}^c\right)+\xi_cN_e\big(\xi_{[a}{T_{b]}}^{ec}-2\xi_{[a}{T^{(e}}_{b]}\,\!^{c)}\big)\nonumber\\
&+\xi_c\xi_e\left(N_fT^{ecf}N_{[a}\xi_{b]}-{T^{ec}}_{[a}N_{b]}\right)
\end{align}
It is much clearer now from this expression that the twist tensor does not indeed vanish in general unless all the terms in expression (\ref{FinalTwist}) do. More specifically, we easily check that the twist tensor simplifies but does not vanish for the special case of a totally antisymmetric torsion either, a case which will be of interest to us below. In fact, for such a special case, the twist tensor (\ref{FinalTwist}) reduces to,
\begin{equation}\label{AntiTwist}
\omega_{ab}=-\tfrac{1}{2}\xi^c\left(T_{cab}+N^e\xi_{b}T_{aec}-N^e\xi_{a}T_{bec}\right).
\end{equation}

Let us now find Raychaudhuri's equation for the geodesics congruence of the null vectors $\xi^a$. Since our goal here is to use such an equation to clearly see the effect of torsion on wormholes, the form of the equation as given recently in Ref.~\cite{LuzVitagliano} cannot be of much use to us. Indeed, we would like to have such an equation written in terms of the torsion tensor and its derivatives, but without the deviation tensor $B_{ab}$. Let us use the usual convention and define the trace of torsion by $T_a={T^b}_{ab}$. In addition, let us keep in mind that $\xi^a\nabla_a\xi^b=0$, $\xi^a\nabla_b\xi_a=0$, $\xi^a\nabla_aN^b=0$\footnote{This can easily be seen by noticing that $N_b\xi^a\nabla_aN^b=0$, implying that $\xi^a\nabla_aN^b=fN^b$ for some scalar $f$. But, since $\xi_b\xi^a\nabla_aN^b=0$, we immediately conclude that $f=0$.}, and that in spacetimes with torsion we have $[\nabla_a,\nabla_b]\xi^d=-{R_{abc}}^d\xi^c-{T^c}_{ab}\nabla_c\xi^d$ \cite{SabbataGasperini}. Then, with the use of Eqs.~(\ref{Expansion}), (\ref{SpatialB}), (\ref{BTensor}) and (\ref{SpatialBFullB}), we find a first form of Raychaudhuri's equation as follows,
\begin{align}\label{BeforeFinalRaychaudhuri}
\tfrac{\rm d}{{\rm d}\lambda}\theta=&\,\xi^a\nabla_a\theta\nonumber\\
=&\,\xi^a\nabla_a\nabla_b\xi^b+\xi^a\nabla_a\left(\xi^cT_c+N^b\xi^c\xi^dT_{cdb}\right)\nonumber\\
=&\,\xi^a\nabla_b\nabla_a\xi^b\!-\!\xi^ a{R_{abd}}^b\xi^d\!-\!\xi^a{T^c}_{ab}\nabla_c\xi^b\!+\!\xi^c\xi^a\nabla_aT_c\nonumber\\
&+\left(\xi^a\nabla_aN^b\right)\xi^c\xi^dT_{cdb}+N^b\xi^c\xi^d\xi^a\nabla_aT_{cdb}\nonumber\\
=&-\nabla_b\xi_a\nabla^a\xi^b-\!R_{ab}\xi^a\xi^b\!-\!\xi_aT^{cab}B_{bc}\!+\!\xi_aT^{cab}\xi^dT_{bdc}\nonumber\\
&+\xi^c\xi^a\nabla_aT_c+N^b\xi^c\xi^d\xi^a\nabla_aT_{cdb}\nonumber\\
=&-B_{ab}B^{ba}-R_{ab}\xi^a\xi^b+\xi_aT^{cab}B_{bc}+\xi^c\xi^a\nabla_aT_c\nonumber\\
&+N^b\xi^c\xi^d\xi^a\nabla_aT_{cdb}.
\end{align}

Our next goal now is to trade the quadratic term $B_{ab}B^{ba}$ in this form of Raychaudhuri's equation for the product $\mathbf{B}_{ab}\mathbf{B}^{ba}$ that will allow us, thanks to Eq.~(\ref{SpatialB}), to recast the equation purely in terms of the expansion, the shear, the twist, the Ricci tensor and torsion. For that purpose, we notice first that $N^a\nabla_a\xi_b=f\xi_b$ for some scalar $f$ because $\xi^bN^a\nabla_a\xi_b=0$. Therefore, using the definition (\ref{BTensor}), we deduce that,
\begin{equation}\label{UsefulIdentities}
N^bB_{ab}=f\xi_a+N^b\xi^cT_{acb}.
\end{equation}
This identity will allows us to get rid of any appearance of $B_{ab}$ --- such as $T^{cab}B_{bc}$ --- in the final Raychaudhuri equation. Indeed, using identity (\ref{UsefulIdentities}) together with Eqs.~(\ref{SpatialBFullB}), (\ref{SpatialBSquared}) and the last equality in Eq.~(\ref{BeforeFinalRaychaudhuri}), we get finally Raychaudhuri's equation in the sought-after pure form as follows,
\begin{align}\label{FinalRaychaudhuri}
\tfrac{\rm d}{{\rm d}\lambda}\theta=&-\tfrac{1}{2}\theta^2-\sigma_{ab}\sigma^{ab}+\omega_{ab}\omega^{ab}-R_{ab}\xi^a\xi^b\nonumber\\
&+\xi_aT^{cab}\left(\sigma_{bc}+\omega_{bc}+\tfrac{1}{2}\theta h_{bc}\right)\nonumber\\
&+\xi^c\xi^a\nabla_aT_c+N^b\xi^c\xi^d\xi^a\nabla_aT_{cdb}.
\end{align}

We now come to the final step that would reveal the effect the nature of matter has on the wormhole by making the energy-momentum tensor appear on the right-hand side of Eq.~(\ref{FinalRaychaudhuri}). For that purpose, we need Einstein-Cartan-Sciama-Kibble metric field equations, which read as follows \cite{SabbataGasperini,HehlReview},
\begin{align}\label{ECSK}
G_{ab}=&\;T^M_{ab}+\tfrac{1}{2}\left(\nabla^c+T^c\right)\left(T_{cab}+T_{acb}+T_{bca}\right)\nonumber\\
&+\left(\nabla^c+T^c\right)\left(g_{ac}T_b-g_{ab}T_c\right).
\end{align}
Contracting both sides of these equations by $\xi^a\xi^b$, we immediately extract the term $R_{ab}\xi^a\xi^b$ in terms of the matter energy-momentum tensor and torsion. Substituting the result inside Raychaudhuri's equation (\ref{FinalRaychaudhuri}), we find,
\begin{align}\label{RaychaudhuriWithMatter}
\tfrac{\rm d}{{\rm d}\lambda}\theta&=-\tfrac{1}{2}\theta^2-\sigma_{ab}\sigma^{ab}+\omega_{ab}\omega^{ab}-T^M_{ab}\xi^a\xi^b\nonumber\\
&+\xi_aT^{cab}\left(\sigma_{bc}+\omega_{bc}+\tfrac{1}{2}\theta h_{bc}\right)-\xi^a\xi^b\left(T^cT_{acb}+T_aT_b\right)\nonumber\\
&-\xi^a\xi^b\left(N^d\xi^c\nabla_cT_{adb}+\nabla^cT_{acb}\right).
\end{align}

Contemplating this expression we immediately notice that the last line contains derivatives of torsion. The sign contribution of such derivatives to the whole equation cannot be determined without knowing the spin matter distribution around the throat. As such, no general conclusion about the fate of the NEC of matter could be reached either, in contrast to the conclusions made in Refs.~\cite{HochbergVisser,Battista}. As a consequence, a case-by-case study should be conducted instead. For this reason, we shall pursue our study by distinguishing four types of spin matter fields that have been well studied in the literature in relation to torsion.
\subsubsection{With scalar, Maxwell and Yang-Mills fields}
A scalar field has spin zero and, thanks to the second set of Einstein-Cartan-Sciama-Kibble field equations that relate the spin-density tensor to torsion  \cite{SabbataGasperini,HehlReview}, the latter also vanishes. In this case, we conclude that scalar fields have to violate the NEC to make a traversable wormhole in agreement with Ref.~\cite{Battista}.

In the case of the Maxwell field, like the gauge Yang-Mills fields, it is known that a minimal coupling of these with torsion violates gauge invariance \cite{HehlReview}. Therefore, because of gauge invariance, Maxwell and Yang-Mills fields are not allowed to minimally couple to torsion and the latter does not consequently arise in the presence of such fields. Thus, these fields also have to violate the NEC to make a traversable wormhole. We keep away from the extra complications of non-minimal coupling with torsion in this paper as such coupling involves some degree of arbitrariness that cannot be of much relevance to our present goal.
\subsubsection{With a fermion field}
In the presence of a fermion field $\psi$ of spin-$\tfrac{1}{2}$ minimally coupled to spacetime torsion, the latter is given by \cite{HehlReview},
\begin{equation}\label{TForFermions}
T^{abc}=\tfrac{1}{4}\bar{\psi}\gamma^{[a}\gamma^b\gamma^{c]}\psi.
\end{equation}
Here, the matrices $\gamma^a$ are the spacetime Dirac matrices related to the constant Dirac matrices $\gamma^i$ through the tetrad fields by $\gamma^a=e^a_i\gamma^i$ \cite{HehlReview}. It is clear from Eq.~(\ref{TForFermions}) that torsion for Dirac fermions is completely antisymmetric. Setting $\theta=0$ and taking a totally antisymmetric torsion, Eq.~(\ref{RaychaudhuriWithMatter}) simplifies greatly and reduces to,
\begin{equation}\label{AntiRaychaudhuriWithMatter}
\tfrac{\rm d}{{\rm d}\lambda}\theta+\sigma_{ab}\sigma^{ab}=\omega_{ab}\omega^{ab}-T^M_{ab}\xi^a\xi^b+\xi_aT^{cab}\omega_{bc}.
\end{equation}
Thus, to have a traversable wormhole in this case, the requirement is that the energy-momentum tensor of the fermions $T^M_{ab}$ satisfies the following inequality,
\begin{equation}\label{InequalityforT}
T^M_{ab}\xi^a\xi^b<\omega_{ab}\omega^{ab}+\xi_aT^{cab}\omega_{bc}.
\end{equation}
Setting $T_{abc}=\epsilon_{abcd}v^d$, for an arbitrary vector $v^a$, and substituting inside expression (\ref{AntiTwist}) of the twist tensor, we find, after using the identities $\epsilon^{abcd}\epsilon_{aefg}=-6\delta^{[b}_e\delta^{c}_f\delta^{d]}_g$ and $\epsilon^{abcd}\epsilon_{abef}=-4\delta^{[c}_e\delta^{d]}_f$ \cite{Wald}, that the right-hand side of inequality (\ref{InequalityforT}) takes the form,
\begin{equation}\label{TomegaTerm}
\omega_{ab}\omega^{ab}+\xi_aT^{cab}\omega_{bc}=\tfrac{1}{2}(\xi_av^a)^2-(\xi_av^a)^2<0.
\end{equation}
Thus, the contraction $T^M_{ab}\xi^a\xi^b$ has to be negative. This shows that making a traversable wormhole with a Dirac field requires the latter to violate the NEC even if one is working in Riemann-Cartan spacetime.
\subsubsection{With a Proca field}
A Proca field is a massive spin-$1$ vector field $A_a$. Introducing the usual tetrad-based notation $e\equiv\sqrt{-g}$, with $g$ being the spacetime metric determinant, the Lagrangian of a Proca field of mass $m$ \cite{HehlReview}, as well as the field equations one extracts from a given Proca Lagrangian \cite{Seitz}, are, respectively,
\begin{align}\label{ProcaLagrangian}
\mathscr{L}=-\tfrac{1}{2}e\big(\nabla_{[a}A_{b]}\nabla^{[a}A^{b]}-m^2A_aA^a\big).\\
\label{ProcaEOM1}
\partial_a\big(e\nabla^{[a}A^{b]}\big)+\tfrac{1}{2}e{T^b}_{ac}\nabla^{[a}A^{c]}+em^2A^b=0.
\end{align}
By using the metricity condition, as well as the two identities, $\partial_a e=eC^b_{ab}$ and $C^b_{ba}=C^b_{ab}-T_a$, which one can deduce using $g^{-1}\partial_a g=-g_{bc}\partial_a g^{bc}$ \cite{Wald} as well as Eqs.~(\ref{TorsionCovariantDerivative1}) and (\ref{TorsionConnection}), the previous field equations can be recast into the following more useful form to us,
\begin{equation}\label{ProcaEOM2}
\big(\nabla_a+T_{a}\big)\nabla^{[a}A^{b]}+m^2A^b=0.
\end{equation}
  On the other hand, the spin-density tensor of a Proca field as presented in Ref.~\cite{HehlReview} gives rise to the following torsion tensor and its trace,
\begin{align}
T_{abc}=&\,A_b\nabla_{[c}A_{a]}-A_c\nabla_{[b}A_{a]}\nonumber\\
&+\tfrac{1}{2}A^d\big(g_{ab}\nabla_{[d}A_{c]}-g_{ac}\nabla_{[d}A_{b]}\big).\label{TForProca}\\
T_{b}=&-\tfrac{1}{2}A^a\nabla_{[a}A_{b]}.\label{TraceForProca}
\end{align}

Let us now compute individually each of the terms in Raychaudhuri's equation (\ref{RaychaudhuriWithMatter}) in terms of the vector field $A^a$ and its derivatives using expressions (\ref{TForProca}) and (\ref{TraceForProca}).
\begin{align}\label{EachTerm}
&\xi_aT^{cab}\sigma_{bc}=-\xi_a\sigma_{bc}A^b\nabla^{[a}A^{c]},\nonumber\\
&\xi_aT^{cab}\omega_{bc}=\xi_aA^a\omega_{bc}\nabla^{[b}A^{c]}-A_b\omega^{bc}\xi^a\nabla_{[a}A_{c]},\nonumber\\
&\xi^a\xi^bT^cT_{acb}=-(\xi_aT^a)^2+m^2(\xi_aA^a)^2+\xi_aA^a\xi_b\nabla_c\nabla^{[c}A^{b]}\!,\nonumber\\
&\xi^a\xi^bN^d\xi^c\nabla_cT_{adb}=-\xi^c\nabla_c\big(\xi^a\xi^bN^dA_b\nabla_{[d}A_{a]}-T_b\xi^b\big),\nonumber\\
&\xi^a\xi^b\nabla^cT_{acb}=-\xi^a\xi^b\big(A_b\nabla^c\nabla_{[c}A_{a]}\!+\!\nabla_{[c}A_{a]}\nabla^cA_b\!+\!\nabla_aT_b\big)\!.
\end{align}
Substituting these inside Eq.~(\ref{RaychaudhuriWithMatter}), and taking into account that $\theta=0$, many terms cancel among themselves and the equation simplifies to the following,
\begin{align}\label{RaychaudhuriProca}
\tfrac{\rm d}{{\rm d}\lambda}\theta+\sigma_{ab}\sigma^{ab}=&\;\omega_{ab}\omega^{ab}-T^M_{ab}\xi^a\xi^b\nonumber\\
&+\!\xi_aA^b(\sigma_{bc}+\omega_{bc})\nabla^{[c}A^{a]}\!+\!\xi_aA^a\omega_{bc}\nabla^{[b}A^{c]}\nonumber\\&-m^2(\xi_aA^a)^2+\xi^a\xi^b\nabla_{[c}A_{a]}\nabla^cA_b\nonumber\\
&+\xi^a\xi^bN^dA_a\xi^c\nabla_c\nabla_{[d}A_{b]}.
\end{align}

We see from this equation that even with the help of the simplified field equations (\ref{ProcaEOM2}) it is still impossible to find the overall sign of all those extra terms on the right-hand side without solving the field equations (\ref{ProcaEOM2}) first. Indeed, only by solving those second order equations of motion for $A^a$ --- by taking into account the boundary conditions of the system --- will one be able to find the various gradients of the field $A^a$ in Eq.~(\ref{RaychaudhuriProca}). This is in complete contrast to what was concluded in Ref.~\cite{Battista} --- without using Raychaudhuri's equation --- about the general non-violation of the NEC by the Proca field to make a traversable wormhole. 

Fortunately, however, the structure of those extra terms still allows one to extract some important conclusions with regard to some special cases even without solving the vector field's equations of motion. In fact, we notice that the majority of those terms contain the contraction $\xi_aA^a$. Therefore, choosing a polarization for the vector $A^a$ such that $\xi_aA^a=0$ will cancel many of the terms in Eq.~(\ref{RaychaudhuriProca}). Such a constraint on the vector field $A^a$ would just mean that either the vector is parallel to the null direction $\xi^a$ or that the vector just has all its components lie in the transverse space. 

Let us start with the case $A^a=f\xi^a$, for some scalar $f$. We immediately see, by recalling that $\xi^a\nabla_a\xi^b=0$ and that both $\sigma_{ab}$ and $\omega_{ab}$ are transverse, that the majority of the extra terms in Eq.~(\ref{RaychaudhuriProca}) vanish and the latter reduces to, 
\begin{equation}\label{SimpleRaychaudhuriProca1}
\tfrac{\rm d}{{\rm d}\lambda}\theta+\sigma_{ab}\sigma^{ab}=\omega_{ab}\omega^{ab}-T^M_{ab}\xi^a\xi^b.
\end{equation}
On the other hand, with $A^a=f\xi^a$ we have $A^a\nabla_bA_a=0$ and $A^a\nabla_aA_b=A_bA^a\nabla_a\ln f$. Using these identities, together with Eq.~(\ref{FinalTwist}), we find that $\omega_{ab}=0$. Therefore, we conclude that in the case of a {\it null} Proca vector field, the requirement for a traversable wormhole in the Riemann-Cartan spacetime is $T^M_{ab}\xi^a\xi^b<0$, i.e., the NEC has to be violated by the field's energy-momentum tensor.

Let us now consider the case of a transverse field. This case implies that we have again the identity $\xi_aA^a=0$. In addition, if we insist that the vector field remains permanently transverse we also have its derivatives constantly lie in the transverse space. This translates into, $(v_a\nabla^aA^b)\xi_b=0$ for any arbitrary vector $v^a$. Implementing these extra conditions inside Eq.~(\ref{RaychaudhuriProca}), many of the extra terms disappear again and we are left with the following equation,
\begin{align}\label{CanceledTermsRaychaudhuriProca}
\tfrac{\rm d}{{\rm d}\lambda}\theta+\sigma_{ab}\sigma^{ab}=&\;\omega_{ab}\omega^{ab}-T^M_{ab}\xi^a\xi^b\nonumber\\
&+\xi_aA^b(\sigma_{bc}+\omega_{bc})\nabla^{[c}A^{a]}.
\end{align}

We are going to show that the first term $\omega_{ab}\omega^{ab}$ on the right-hand side of this equation does not vanish while the sum in the second line of the equation does. Let us start with the latter. For $\theta=0$, which is the case in Eq.~(\ref{CanceledTermsRaychaudhuriProca}), we have $\sigma_{ab}+\omega_{ab}=\mathbf{B}_{ab}$. Then, using Eqs.~(\ref{BTensor}) and (\ref{SpatialBFullB}), we have the following, 
\begin{align}\label{IntermediateCanceledTermsRaychaudhuriProca}
\xi_aA^b\mathbf{B}_{bc}\nabla^{[c}A^{a]}=&-\xi_aA^b\mathbf{B}_{bc}\nabla^{a}A^{c}\nonumber\\
=&-\xi_aA^b(\nabla_c\xi_b+\xi^eT_{bec})\nabla^{a}A^{c}\nonumber\\
&-\xi_aA^bN^e\left(\xi_cB_{be}+\xi_bB_{ec}\right)\nabla^{a}A^{c}\nonumber\\
&-\xi_aA^bN_b\xi^e\xi^fT_{efc}\nabla^{a}A^{c}\nonumber\\
&-\xi_aA^bN^d\xi_b\xi_cN^eB_{ed}\nabla^{a}A^{c}\nonumber\\
&-\xi_aA^bN^dN_b\xi_c\xi^e\xi^fT_{efd}\nabla^{a}A^{c}.
\end{align}
In the first line we have used the identity $(v_a\nabla^aA^b)\xi_b=0$. Now, with a repeated use of this identity, we easily see that each term of the subsequent two lines in Eq.~(\ref{IntermediateCanceledTermsRaychaudhuriProca}) vanishes. In addition, each term of the three last lines vanishes also because of the fact that $A^a$ lives in the transverse space. The only term for which it is not straightforward to see why it vanishes is the term $\xi_aA^b\xi^eT_{bec}\nabla^{a}A^{c}$. We can check that this term is zero by substituting expression (\ref{TForProca}) for the torsion tensor and then performing the contractions by, in addition, taking into account the fact that $\xi_a\nabla^a(A_bA^b)=0$. This condition is necessary for the magnitude of the field $A^a$ to always remain bounded along the null direction $\xi^a$.
Thus, we conclude that Eq.~(\ref{CanceledTermsRaychaudhuriProca}) reduces to Eq.~(\ref{SimpleRaychaudhuriProca1}). Let us then finally check the fate of the twist tensor $\omega_{ab}$ for this special case of a purely transverse field $A^a$.

Substituting the expression (\ref{TForProca}) of torsion inside Eq.~(\ref{FinalTwist}) and then taking into account the two identities $\xi_aA^a=0$ and $(v_a\nabla^aA^b)\xi_b=0$, we end up with,  
\begin{align}\label{TransverseProcaTwist}
\omega_{ab}=\tfrac{1}{2}A_c(\xi_{[a}\nabla^cA_{b]}-\xi_{[a}\nabla_{b]}A^c).
\end{align}
Without any additional constraint on the vector field $A^a$, the right-hand side of this identity does not vanish in general. 
Given that $\omega_{ab}$ is, in addition, purely spatial, we conclude that $\omega_{ab}\omega^{ab}>0$. Thus, the requirement for having a traversable wormhole in this case reduces only to $T^M_{ab}\xi^a\xi^b<\omega_{ab}\omega^{ab}$ and the energy-momentum tensor of a {\it transverse} Proca field does not have to violate the NEC.
\subsubsection{With a Rarita-Schwinger field}
A Rarita-Schwinger field $\psi_a$ is a spin-$\tfrac{3}{2}$ vector-spinor field --- that we take to be massless for simplicity --- the spin-density tensor of which creates a torsion tensor given by \cite{Hehl}\footnote{We use here the normalization conventions of Ref.~\cite{Nieuwenhuizen}.},
\begin{equation}\label{TForRS}
{T^c}_{ab}=\tfrac{1}{2}\bar{\psi}_a\gamma^c\psi_b.
\end{equation}
This field is, in addition, constrained to satisfy the condition $\gamma^a\psi_a=0$ \cite{Nieuwenhuizen}. This immediately yields a traceless torsion since $T_a={T^b}_{ab}=0$. Thanks to this fact and setting $\theta=0$, Eq.~(\ref{RaychaudhuriWithMatter}) simplifies to,
\begin{align}\label{RaychaudhuriRS}
\tfrac{\rm d}{{\rm d}\lambda}\theta+\sigma_{ab}\sigma^{ab}=&\;\omega_{ab}\omega^{ab}-T^M_{ab}\xi^a\xi^b+\xi_aT^{cab}\left(\sigma_{bc}+\omega_{bc}\right)\nonumber\\
&-\xi^a\xi^b\left(N^d\xi^c\nabla_cT_{adb}+\nabla^cT_{acb}\right).
\end{align}

As for the case of the Proca field, the subsistence of the torsion gradients in this equation makes it impossible to conclude anything about the sign contribution of these gradients without having their values. In fact, although the equation of motion for the field $\psi_a$ is first order and reads, $\gamma^a(D_a\psi_a-D_b\psi_a)=0$ \cite{Nieuwenhuizen}, where $D_a$ is the covariant derivative operator that acts on spinors, these equations cannot be of any help without solving them first. This is due to the various contractions with the null vector, like $\xi_aT^{cab}=\tfrac{1}{2}\xi_a\bar{\psi}^a\gamma^c\psi^b$ and the derivatives of $\xi^bT_{acb}=\tfrac{1}{2}\xi^b\bar{\psi}_c\gamma_a\psi_b$, the values of which depend on the specific form of the field $\psi_a$. Nevertheless, the specific structure of these terms that hinders us from reaching any conclusion suggests the possibility of treating instead the special case of $\xi^a\psi_a=0$. This, like for the Proca field, just means that the vector-spinor field is chosen either {\it parallel} or {\it transverse} to the null direction. As we shall see now, the conclusion one gets for this special case does not depend on which of these two possibilities one chooses.

By implementing this restriction inside Eq.~(\ref{RaychaudhuriRS}), all the extra terms --- apart from the first one --- on the right-hand side disappear and we are left with Eq.~(\ref{SimpleRaychaudhuriProca1}). All we have to do then is to find the value of the twist tensor. Plugging expression (\ref{TForRS}) inside Eq.~(\ref{FinalTwist}) and taking into account the constraint $\gamma^a\psi_a=0$, we find,
\begin{equation}\label{TwistForRS}
\omega_{ab}=\tfrac{1}{4}\bar{\psi}_a\xi_c\gamma^c\psi_b,
\end{equation}
which is clearly non-vanishing without any further constraint on $\psi_a$. As before, given the purely spatial nature of the twist tensor, we conclude that $\omega_{ab}\omega^{ab}>0$. Hence the energy-momentum tensor of the spin-$\tfrac{3}{2}$ field under this special constraint does not have to violate the NEC as all that is required to have a traversable wormhole is $T^M_{ab}\xi^a\xi^b<\omega_{ab}\omega^{ab}$.
\section{In the conformal frame}\label{sec:3}
\subsection{Without torsion}

To find Raychaudhuri's equation in the conformal frame one either finds first the conformal transformation of each of the quantities (shear, twist, null vector, and Rici tensor) inside that equation and then proceeds with the usual derivation of the equation based on these new quantities, or just notices that the equation is purely geometric in nature and therefore automatically remains invariant under the conformal transformation \cite{AugustPaper}. This means that all one has to do to write down the conformally transformed version is decorate every symbol inside the original equation (\ref{TorsionFreeRaychau}) by tildes. The resulting equation is thus,
\begin{equation}\label{ConfRaychau}
\tfrac{\rm d}{{\rm d}\tilde{\lambda}}\tilde{\theta}=-\tfrac{1}{2}\tilde{\theta}^2-\tilde{\sigma}_{ab}\tilde{\sigma}^{ab}+\tilde{\omega}_{ab}\tilde{\omega}^{ab}-\tilde{R}_{ab}\tilde{\xi}^a\tilde{\xi}^b.
\end{equation}

The next step is to notice that the null vector $\tilde{\xi}^a$ of the conformal frame remains hypersurface orthogonal, such that, based on Frobenius' theorem for torsion-free spacetimes, we also have $\tilde{\omega}_{ab}=0$. In addition, the shear tensor $\tilde{\sigma}_{ab}$ remains spacelike, thereby making $\tilde{\sigma}_{ab}\tilde{\sigma}^{ab}$ positive everywhere. On the other hand, recall that under a Weyl transformation Einstein equations transform as follows \cite{DabrowskiAl},
\begin{equation}\label{ConfEinsteinEqs}
\tilde{G}_{ab}=T^M_{ab}-\frac{2\tilde{\nabla}_a\tilde{\nabla}_b\Omega}{\Omega}+\tilde{g}_{ab}\left[\frac{2\tilde{\Box}\Omega}{\Omega}-\frac{3\tilde{\nabla}_c\Omega\tilde{\nabla}^c\Omega}{\Omega^2}\right].
\end{equation}
After contracting both sides of these equations by $\tilde{\xi}^a\tilde{\xi}^b$ we obtain the desired contraction $-\tilde{R}_{ab}\xi^a\xi^b$ in terms of the energy-momentum tensor $T^M_{ab}$. Substituting the resulting expression inside Eq.~(\ref{ConfRaychau}), and then recalling that $T^M_{ab}=\Omega^2 \tilde{T}^M_{ab}$ \cite{DabrowskiAl}, we find,  
\begin{equation}\label{BeforeConfNEC}
\tfrac{\rm d}{{\rm d}\tilde{\lambda}}\tilde{\theta}+\tilde{\sigma}_{ab}\tilde{\sigma}^{ab}=-\Omega^2\tilde{T}^M_{ab}\tilde{\xi}^a\tilde{\xi}^b+\frac{2\tilde{\xi}^a\tilde{\xi}^b\tilde{\nabla}_a\tilde{\nabla}_b\Omega}{\Omega}.
\end{equation}
We clearly see now that the energy-momentum tensor of matter $\tilde{T}^M_{ab}$, as perceived from the conformal frame, does not necessarily have to violate the null energy condition, but is only required to satisfy the following inequality,
\begin{equation}\label{ConfNEC}
\tilde{T}^M_{ab}\tilde{\xi}^a\tilde{\xi}^b<\frac{2\tilde{\xi}^a\tilde{\xi}^b\tilde{\nabla}_a\tilde{\nabla}_b\Omega}{\Omega^3},
\end{equation}
with the right-hand side not being necessarily negative\footnote{Note that in Ref.~\cite{ValerioAl}, where the nature of the four Brans solution classes of Brans-Dicke theory have been studied in detail, a field redefinition has been performed after applying the conformal transformation. That field redefinition was made to obtain a simple expression for the scalar field's energy-momentum tensor on the right-hand side of Einstein equations. For all four classes, the field redefinitions displayed there would make the NEC satisfied in the conformal frame despite the existence of Brans wormholes in the new frame. In addition, such field redefinitions induced a disagreement with the sign of the Einstein tensor, and hence of the Ricci scalar, discussed for class IV solution recently in Ref.~\cite{BoonsermAl} (F.H. is grateful to Valerio Faraoni for bringing this reference to his attention which led to the discussion in this footnote). These two disagreements are actually simply due to a missing factor of the pure imaginary $i$ in the field redefinitions (2.26), (3.20), (4.26) and (5.15) given in Ref.~\cite{ValerioAl}. Putting this factor back, one easily recovers the violation of the NEC and the missing negative ($-$) sign in the Ricci scalars given for each of the corresponding solution classes there.}. 

Let us examine a simple example. Take the conformally transformed tidal force-free Morris-Thorne wormhole metric \cite{MorrisThorne,HochbergVisser},
\begin{equation}\label{ConfMTWormhole}
{\rm d}\tilde{s}^2=\Omega^2(t)\left(\!-dt^2+\frac{{\rm d}r^2}{1-b(r)/r}+r^2{\rm d}\vartheta^2+r^2\sin^2\!\vartheta{\rm d}\phi^2\right)\!\!.
\end{equation}
Here, $b(r)$ is a function such that $b(r_0)=r_0$ at the location $r_0$ of the throat \cite{MorrisThorne}. We chose here the conformal factor $\Omega(t)$ to depend only on time for simplicity. With this conformal metric $\tilde{g}_{ab}$, we easily find a null vector $\tilde{\xi}^a$ as follows \cite{HochbergVisser},
\begin{equation}\label{XiForMT}
\tilde{\xi}^a=(\sqrt{2}\Omega)^{-1}\left(1,\sqrt{1-b(r)/r},0,0\right).
\end{equation}
In addition, with the metric (\ref{ConfMTWormhole}) we can also easily compute the Christoffel symbols $\tilde{\Gamma}_{ab}^c$ to be used inside the covariant derivatives $\tilde{\nabla}_a$. The non-vanishing ones that are of interest to us here are, $\tilde{\Gamma}_{00}^0=\dot{\Omega}/\Omega$ and $\tilde{\Gamma}_{11}^0=\dot{\Omega}/[\Omega(1-b(r)/r)]$. With these at hand, we obtain,
\begin{equation}\label{ConfNECWithMT}
\tilde{\xi}^a\tilde{\xi}^b\tilde{\nabla}_a\tilde{\nabla}_b\Omega=\frac{1}{2\Omega^2}\left(\ddot{\Omega}-\frac{2\dot{\Omega}^2}{\Omega}\right).
\end{equation}
An example of a conformal factor that makes the content of the parentheses positive is $\Omega(t)=t^{-1/3}$ (a wormhole embedded in a contracting universe.) 

This shows that the right-hand side of inequality (\ref{ConfNEC}) does not indeed necessarily have to be negative. We conclude that matter does not necessarily have to violate the NEC in the conformal frame either. This confirms again our hand-waving argument presented in the Introduction.
\subsection{With torsion}
To investigate the effect of torsion in the conformal frame, we need to also specify the conformal transformation of torsion. It turns out, however, that because torsion is a separate degree of freedom of spacetime in addition to the metric, there is not one, but three possibilities proposed in the literature for how torsion might transform under a Weyl mapping (see Ref.~\cite{Shapiro} and references therein). The three possible ways are the following,
\begin{align}
{\tilde{T}^c}_{ab}&={T^c}_{ab},\label{ThreeWayTorsion1}\\
{\tilde{T}^c}_{ab}&={T^c}_{ab}+\alpha\left(\delta^c_a\partial_b-\delta^c_b\partial_a\right)\ln\Omega,\label{ThreeWayTorsion2}\\
\tilde{T}_a&=T_a+\alpha\,\partial_a\ln\Omega.\label{ThreeWayTorsion3}
\end{align}
Here, $\alpha$ is some arbitrary parameter \cite{Shapiro}\footnote{These transformations are given in Ref.~\cite{Shapiro} with reference to a non-minimal coupling between a scalar field and torsion. A derivation of these transformations based on the conformal transformation of the tetrads, $e^a_i=\Omega^{-1} e^a_i$, and a conformal invariance of the Lorentz connection ${\varpi^a}_{ij}$, which are both fundamental gauge fields of Einstein-Cartan gravity, is given in Ref.~\cite{Yoon}}. 

On the other hand, the expression (\ref{FinalTwist}) of the twist tensor having been obtained from purely geometric definitions, and the derivations being based on purely geometric steps, we know that the conformally transformed twist tensor $\tilde{\omega}_{ab}$ will keep the same form as in Eq.~(\ref{FinalTwist}), only decorated everywhere with tildes. Similarly, Raychaudhuri's equation for Riemann-Cartan spacetimes in the form (\ref{FinalRaychaudhuri}), being also a purely geometric equality, implies that the conformally transformed version keeps the same form, only decorated everywhere with tildes. As a consequence of this, we see that, just as in the original frame, the presence of the divergence term $\tilde{\nabla}^a\tilde{T}_{bac}$ would not allow one to conclude anything about the NEC without knowing the dynamics of the matter spin in the conformal frame. 

To be able to easily see the effect of torsion on the fate of the NEC in the conformal frame, we should therefore make again the simplifying assumption of a totally antisymmetric torsion tensor in the new frame as well. In this case, however, only the first possibility (\ref{ThreeWayTorsion1}) for the transformation of torsion would be meaningful as the other two would not allow for total antisymmetry of ${\tilde{T}^c}_{ab}$. Implementing this option inside expression (\ref{FinalTwist}), we find the same form for the twist tensor as that of Eq.~({\ref{AntiTwist}), but decorated of course with tildes. 

Next, to find the equivalent of Eq.~(\ref{RaychaudhuriWithMatter}) --- involving matter via the energy-momentum tensor --- we need the conformal transformation of Einstein-Cartan-Sciama-Kibble metric field equations. However, because the Riemann tensor contains products of the form $\tilde{C}^c_{ab}\tilde{C}^a_{cd}$ and that the Christoffel symbols inside $\tilde{C}^c_{ab}$ are affected by the transformation, the field equations will acquire more additional terms on the right-hand side than what Einstein equations do after a conformal transformation even in the case of an invariant torsion. These extra terms would arise from the mixed products of the form $\tilde{\Gamma}^c_{ab}{T^a}_{cd}$. 

Fortunately, all we actually need from the field equations is to extract the contraction $\tilde{R}_{ab}\tilde{\xi}^a\tilde{\xi}^b$ and express it in terms of $T^M_{ab}$. For this purpose, it turns out that a simple inspection would reveal the transformation of such a term. Indeed, the full Ricci tensor in the conformally transformed Riemann-Cartan spacetime reads \cite{SabbataGasperini,HehlReview},
\begin{equation}\label{RiemannInRiemannCartan}
{\tilde{R}^c}_{acb}=\partial_c\tilde{C}^c_{ab}-\partial_b\tilde{C}^c_{ac}+\tilde{C}^c_{ce}\tilde{C}^e_{ab}-\tilde{C}^c_{be}\tilde{C}^e_{ca}.
\end{equation}
To obtain the term $\tilde{R}_{ab}\tilde{\xi}^a\tilde{\xi}^b$ we would have to insert the conformal expression of each of the connection terms and then contract both sides of this identity by $\tilde{\xi}^a\tilde{\xi}^b$. However, given that we work with a totally antisymmetric torsion and that $\tilde{C}^c_{ab}=\tilde{\Gamma}^c_{ab}+\tfrac{1}{2}\big({\tilde{T}^c}_{ab}+\tilde{T}_a\,^c\,_b+\tilde{T}_b\,^c\,_a\big)$, we immediately see that the transformed $\tilde{R}_{ab}\tilde{\xi}^a\tilde{\xi}^b$ would involve only exactly the same terms arising when performing a conformal transformation in torsion-free spacetimes. Thus, Raychaudhuri's equation in the conformal frame with matter and totally antisymmetric torsion, takes the same form as Eq.~(\ref{AntiRaychaudhuriWithMatter}) with one extra term borrowed from Eq.~(\ref{BeforeConfNEC}):
\begin{align}\label{ConfAntiRaychaudhuriWithMatter}
\tfrac{\rm d}{{\rm d}\tilde{\lambda}}\tilde{\theta}+\tilde{\sigma}_{ab}\tilde{\sigma}^{ab}=&\;\tilde{\omega}_{ab}\tilde{\omega}^{ab}-\Omega^2\tilde{T}^M_{ab}\tilde{\xi}^a\tilde{\xi}^b+\tilde{\xi}_a\tilde{T}^{cab}\tilde{\omega}_{bc}\nonumber\\
&+\frac{2\tilde{\xi}^a\tilde{\xi}^b\tilde{\nabla}_a\tilde{\nabla}_b\Omega}{\Omega}.
\end{align}
Therefore, all that is required of matter in the conformal frame in order to have a traversable wormhole is to satisfy the following inequality,
\begin{equation}\label{ConfTorsionNEC}
\tilde{T}^M_{ab}\tilde{\xi}^a\tilde{\xi}^b<\frac{\tilde{\omega}_{ab}\tilde{\omega}^{ab}}{\Omega^2}+\frac{\tilde{\xi}_a\tilde{T}^{cab}\tilde{\omega}_{bc}}{\Omega^2}
+\frac{2\tilde{\xi}^a\tilde{\xi}^b\tilde{\nabla}_a\tilde{\nabla}_b\Omega}{\Omega^3}.
\end{equation}
This just shows again that the energy-momentum tensor of matter, as perceived from the conformal Riemann-Cartan spacetime, does not necessarily have to violate the NEC to sustain a traversable wormhole in such a spacetime.
\section{Summary \& Discussion}\label{Conclusion}
The fate of the violation of the NEC required by traversable wormholes has been found to be affected both by adding torsion to spacetime and by performing a Weyl conformal transformation on the latter. We learned, though, that no {\it general} rule could be drawn for the effect of torsion on the NEC because a case-by-case study is needed to figure out the sign contribution of the torsion terms to Raychaudhuri's equation. Nevertheless, we found special cases of matter configurations for which a definite answer could be provided. 

We found that for a Dirac field, for which spacetime torsion is a totally antisymmetric tensor, the NEC still has to be violated to make a traversable wormhole. The required degree of violation of the NEC for this case is found to be enhanced compared to what is required in the absence of torsion. It is worth noting that such a violation has actually emerged merely from the total antisymmetry of the spacetime torsion, and did not depend in any case on the specific value of the torsion generated by the Dirac field. In other words, when the tensorial piece of torsion is absent and only the vectorial piece is present, one is guaranteed to have the NEC violated by traversable wormholes. As such, our result rather suggests the following more general conclusion: In spacetimes with a totally antisymmetric torsion tensor, traversable wormholes are bound to greatly violate the NEC regardless of the specific nature of the matter that creates such torsion and such wormholes.

In contrast, we found that with Proca and Rarita-Schwinger fields, subject to specific constraints, torsion, as with conformal transformations, helps remove completely the requirement to have exotic matter to sustain a traversable wormhole. The needed ``exoticity", supposed to be carried by matter in Riemannian spacetimes, becomes available within the geometry of the spacetime itself in both cases. In the case of Riemann-Cartan spacetimes, torsion induces an effective energy-momentum tensor that might mimic that of an exotic matter. In the case of a Riemannian spacetime built by conformally transforming another Riemannian spacetime the exotic matter is provided by the induced energy-momentum tensor arising from the deformation brought to the metric. 

In addition, we found that the conformal transformation has the same effect also in Riemann-Cartan spacetimes. However, in this case, as in the original frame, we showed that only for the special case of a totally antisymmetric torsion tensor can one clearly see how geometry is always able to provide the required level of exoticity. This is due to the torsion derivatives involved in Raychaudhuri's equation as well as the products of Christoffel symbols and torsion inside the Riemann tensor, making the conformal transformation of the latter much more involved than that in torsion-free spacetimes.

In this work we have dealt solely with Einstein-Cartan gravity. It is well known, however, that the latter theory of gravity, together with general relativity, are just dynamical degenerate cases of the more general Poincar\'e gauge theory of gravity (see, e.g., Ref.~\cite{Obukhov}). It is therefore interesting to investigate whether the results we found here would still hold in such a more general theory of gravity with torsion. Such an investigation would not be trivial, though. In fact, as can be seen from the field equations (\ref{ECSK}), used to find Raychaudhuri's equation (\ref{RaychaudhuriWithMatter}), even within a theory linear in torsion, nonlinear terms in torsion pop up in the field equations. Notwithstanding this computational limitation, our present work is general enough to allow us to look at the special case of totally antisymmetric torsion even within such a general theory. Indeed, Raychaudhuri's equation, in its purely geometric form (\ref{FinalRaychaudhuri}), is independent of the underlying field equations of the background spacetime. This makes the equation then valid in any Poincar\'e gauge theory of gravity. For a totally antisymmetric torsion, however, Eq.~(\ref{FinalRaychaudhuri}) always reduces to Eq.~(\ref{AntiRaychaudhuriWithMatter}). The latter, in turn, always leads to Eqs.~(\ref{InequalityforT}) and (\ref{TomegaTerm}) for a totally antisymmetric torsion, regardless of the field equations of the theory. It is then clear that the NEC would always have to be violated for such a specific torsion even in the more general Poincar\'e gauge theory of gravity.

Another interesting side of this work is that it shows how much geometry can become involved in the interpretations we make of phenomena displaying the interaction between matter and spacetime. Indeed, a great similarity between the effects provided by torsion on the (non-)violation of the NEC by matter and those induced by a conformal transformation have been revealed. In this sense, if one is willing to consider torsion and its effect on physical phenomena as real, one is also bound to consider as real the effect a conformal transformation has on the way physical phenomena would be perceived in the conformal frame\footnote{For completeness, see Refs.~\cite{Hammad2012,DeyAl,ChakrabortyDey} for an investigation on the relation between torsion and spacetime thermodynamics.}. This conclusion does not, however, constitute a definitive solution to the issue of the physical (non-)equivalence of the two frames. A more specialized and rigorous study of this issue will be presented elsewhere (see, however, Ref.~\cite{ConformalIssue}). 

Nevertheless, this conclusion supports at least what has already been found concerning the non-trivial fate of wormholes and black holes in the conformal frame in Refs.~\cite{FaraoniPrainZambrano,Hammad3}. This concordance just speaks once more in favor of viewing Weyl conformal transformations as a non-trivial tool for learning about the deep connection between geometry and matter. As already shown in Refs.~\citep{Hammad1,Hammad2}, although geometry itself could mimic energy, or equivalently mass, under a conformal transformation the true geometric nature of such a quantity always reveals itself as distinct from that of pure matter. In this work, an analogous, but reversed, scenario came out. What was an exotic feature only to be carried by matter to make a traversable wormhole has been transferred into pure geometry thanks to either ({\it i}\,) additional geometric degrees of freedom already stored in Riemann-Cartan spacetimes or ({\it ii}\,) a deformation brought --- \`a la Weyl --- to the spacetime.  
\begin{acknowledgments}
F.H. gratefully acknowledges the support of the Natural Sciences \& Engineering Research Council of Canada (NSERC) (Grant No.~2017-05388) as well as the STAR Research Cluster of  Bishop's University. \'E.M. is partly supported by NSERC's Undergraduate Student Research Awards (USRA Grant No.~2018-523478). P.L. gratefully acknowledges support from the Fonds de Recherche du Qu\'ebec - Nature et Technologies (FRQNT) via a grant from the Programme de Recherches pour les Chercheurs de Coll\`ege and from  the STAR Research Cluster of  Bishop's University. The authors are grateful to the anonymous referee for his/her valuable comments.
\end{acknowledgments}


\begin{thebibliography}{}
\bibitem{ER} A. Einstein and N. Rosen, ``The particle problem in the general theory of relativity", \href{https://journals.aps.org/pr/abstract/10.1103/PhysRev.48.73}{Phys. Rev. {\bf48}, 73 (1935)}.

\bibitem{MorrisThorne} M.S. Morris and K.S. Thorne, ``Wormholes in spacetime
and their use for interstellar travel: A tool for teaching general relativity'', \href{https://aapt.scitation.org/doi/10.1119/1.15620}{Am. J. Phys. {\bf56}, 395 (1988)}.

\bibitem{Visser} M. Visser, ``{\it Lorentzian Wormholes: From Einstein to Hawking}'' (Springer-Verlag, New York, 1996).

\bibitem{Lobo} F.S.N. Lobo, {\it Classical and Quantum Gravity Research} (Nova Science Publishers, New York, 2008), p.1 [\href{https://arxiv.org/abs/0710.4474}{arXiv:0710.4474}].

\bibitem{HochbergVisser} D. Hochberg and M. Visser, ``Dynamic wormholes, antitrapped
surfaces, and energy conditions'' \href{https://journals.aps.org/prd/abstract/10.1103/PhysRevD.58.044021}{Phys. Rev. D
{\bf58}, 044021 (1998)} [\href{https://arxiv.org/abs/gr-qc/9802046}{arXiv:gr-qc/9802046}].

\bibitem{Hehl}  F.W. Hehl, P. von der Heyde and G.D. Kerlick, ``General relativity with spin and torsion and its deviations from Einstein's theory", \href{https://journals.aps.org/prd/abstract/10.1103/PhysRevD.10.1066}{Phys. Rev. D {\bf10}, 1066 (1974)}.

\bibitem{Kopczy'nski1972} W. Kopczy´nski, ``A non-singular universe with torsion", \href{https://www.sciencedirect.com/science/article/pii/0375960172907141}{Phys. Lett. A {\bf39}, 219 (1972)}. 

\bibitem{Kopczy'nski1973} W. Kopczy´nski, ``An anisotropic universe with torsion", \href{https://www.sciencedirect.com/science/article/pii/037596017390546X}{Phys. Lett. A {\bf43}, 63 (1973)}.

\bibitem{Trautman} A. Trautman, ``Spin and Torsion May Avert Gravitational Singularities", \href{https://link.springer.com/article/10.1038\%2Fphysci242007a0}{Nature {\bf242}, 7 (1973)}.

\bibitem{Tafel1973} J. Tafel, ``A non-singular homogeneous universe with torsion", \href{https://www.sciencedirect.com/science/article/pii/0375960173901084}{Phys. Lett. A {\bf45}, 341 (1973)}.

\bibitem{Tafel1975} J. Tafel, ``Class of cosmological models with torsion and spin", \href{http://www.actaphys.uj.edu.pl/fulltext?series=Reg&vol=6&page=537}{Acta Phys. Pol. B {\bf6}, 537 (1975)}.

\bibitem{Kuchowicz1975A} B. Kuchowicz, ``Cosmology with spin and torsion. Part I. Physical and mathematical foundations", \href{http://adsabs.harvard.edu/abs/1975AcC.....3..109K}{Acta Cosmol. {\bf3}, 109 (1975)}. 

\bibitem{Kuchowicz1975B} B. Kuchowicz, ``Cosmology with spin and torsion. Part II. Spatially homogeneous aligned spin models with the Weyssenhoff fluid", \href{http://adsabs.harvard.edu/abs/1976AcC.....4...67K}{Acta Cosmol. {\bf4}, 67 (1975)}.

\bibitem{Kuchowicz1975C} B. Kuchowicz, ``The Einstein-Cartan equations in astrophysically interesting situations. I. The case of spherical symmetry", \href{http://www.actaphys.uj.edu.pl/fulltext?series=Reg&vol=6&page=555}{Acta Phys. Pol. B {\bf6}, 555 (1975)}.

\bibitem{Kuchowicz1976A} B. Kuchowicz, ``Some cosmological models with spin and torsion. I", \href{https://link.springer.com/article/10.1007/BF00640517}{Astrophys. Space Sci. {\bf 39}, 157 (1976)}.

\bibitem{Kuchowicz1976B} B. Kuchowicz, ``Some cosmological models with spin and torsion. II - Axially symmetric models with a uniform magnetic field", Astrophys. \href{https://link.springer.com/article/10.1007/BF00651197}{Space Sci. {\bf40}, 167 (1976)}.

\bibitem{Kuchowicz1978} B. Kuchowicz, ``Friedmann-like cosmological models without singularity", \href{https://link.springer.com/article/10.1007\%2FBF00759545}{Gen. Relativ. Gravit. {\bf9}, 511 (1978)}.

\bibitem{Gasperini} M. Gasperini, ``Repulsive gravity in the very early universe", \href{https://link.springer.com/article/10.1023\%2FA\%3A1026606925857}{Gen. Rel. Grav. {\bf30}, 1703 (1998)} [\href{https://arxiv.org/abs/gr-qc/9805060}{arXiv:gr-qc/9805060}].

\bibitem{Poplawski} N. Poplawski, ``Nonsingular, big-bounce cosmology from spinor-torsion coupling", \href{https://journals.aps.org/prd/abstract/10.1103/PhysRevD.85.107502}{Phys. Rev. D {\bf85}, 107502 (2012)} [\href{https://arxiv.org/abs/1111.4595}{arXiv:1111.4595}].

\bibitem{Anchordoqui} L.A. Anchordoqui, ``Wormholes in spacetime with torsion", \href{https://www.worldscientific.com/doi/abs/10.1142/S0217732398001169}{Mod. Phys. Lett. A {\bf13}, 1095 (1998)} [\href{https://arxiv.org/abs/gr-qc/9612056}{arXiv:gr-qc/9612056}].

\bibitem{JawadRani} A. Jawad and S. Rani, ``Non-minimal Coupling of Torsion-matter Satisfying Null Energy Condition for Wormhole Solutions", \href{https://link.springer.com/article/10.1140\%2Fepjc\%2Fs10052-016-4560-4}{Eur. Phys. J. C {\bf76}, 704 (2016)} [\href{https://arxiv.org/abs/1612.02042}{arXiv:1612.02042 }].

\bibitem{Battista} E. Di Grezia, E. Battista, M. Manfredonia and G. Miele, ``Spin, torsion and violation of null energy condition in traversable wormholes", \href{https://link.springer.com/article/10.1140\%2Fepjp\%2Fi2017-11799-6}{Eur. Phys. J. Plus {\bf132}, 537 (2017)} [\href{https://arxiv.org/abs/1707.01508}{arXiv:1707.01508}].

\bibitem{SabbataGasperini} V. de Sabbata and M. Gasperini, {\it Introduction to Gravitation} (World Scientific Publishing, Singapore, 1985).

\bibitem{HehlReview} F.W. Hehl, P. von der Heyde, G.D. Kerlick and J.M. Nester, ``General relativity with spin and torsion: Foundations and prospects", \href{https://journals.aps.org/rmp/abstract/10.1103/RevModPhys.48.393}{Rev. Mod. Phys. {\bf48}, 393 (1976)}.

\bibitem{DabrowskiAl} M.P. D\c{a}browski, J. Garecki and D.B. Blaschke, ``Conformal transformations and conformal invariance in gravitation", \href{https://onlinelibrary.wiley.com/doi/abs/10.1002/andp.200810331}{Ann. Phys. (Berlin) {\bf18}, 13 (2009)} [\href{https://arxiv.org/abs/0806.2683}{arXiv:0806.2683}].


\bibitem{Wald} R. Wald, \textit{General Relativity} (Chicago University Press, Chicago, 1984).

\bibitem{Poisson} E. Poisson, \textit{A Relativist's Toolkit} (Cambridge University Press, Cambridge, 2004).

\bibitem{DeyAl} R. Dey, S. Liberati and D. Pranzetti, ``Spacetime thermodynamics in the presence of torsion", 
\href{https://journals.aps.org/prd/abstract/10.1103/PhysRevD.96.124032}{Phys. Rev. D {\bf96}, 124032 (2017)} [\href{https://arxiv.org/abs/1709.04031}{arXiv:1709.04031}].

\bibitem{LuzVitagliano} P. Luz and V. Vitagliano, ``Raychaudhuri equation in space-times with torsion", \href{https://journals.aps.org/prd/abstract/10.1103/PhysRevD.96.024021}{
Phys. Rev. D{\bf96}, 024021 (2017)} [\href{https://arxiv.org/abs/1709.07261}{arXiv:1709.07261}].

\bibitem{Seitz} M. Seitz, ``Proca field in a spacetime with curvature and torsion", \href{http://iopscience.iop.org/article/10.1088/0264-9381/3/6/023/meta}{Class. Quantum Grav. {\bf3}, 1265 (1986)}.

\bibitem{Nieuwenhuizen} P. van Nieuwenhuizen, ``Supergravity", \href{https://www.sciencedirect.com/science/article/pii/0370157381901575}{
Phys. Rept. {\bf68}, 189 (1981)}.

\bibitem{AugustPaper} F. Hammad, \'E. Mass\'e and P. Labelle, ``Black hole mechanics and thermodynamics in the light of Weyl transformations", \href{https://journals.aps.org/prd/abstract/10.1103/PhysRevD.98.104049}{Phys. Rev. D {\bf98}, 104049 (2018)} [\href{https://arxiv.org/abs/1811.12201}{arXiv:1811.12201}]. 

\bibitem{ValerioAl} V. Faraoni, F. Hammad and S.D. Belknap-Keet, ``Revisiting the Brans solutions of scalar-tensor gravity", \href{https://journals.aps.org/prd/abstract/10.1103/PhysRevD.94.104019}{Phys. Rev. D {\bf94}, 104019 (2016)} [\href{https://arxiv.org/abs/1609.02783}{arXiv:1609.02783}].

\bibitem{BoonsermAl} P. Boonserm, T. Ngampitipan, A. Simpson and M. Visser, ``The exponential metric represents a traversable wormhole", [\href{https://arxiv.org/abs/1805.03781}{arXiv:1805.03781}].

\bibitem{Shapiro} P. Shapiro, ``Physical Aspects of the Space-Time Torsion", 	\href{https://www.sciencedirect.com/science/article/pii/S0370157301000308?via\%3Dihub}{Phys. Rept. {\bf357}, 113 (2002)} [\href{https://arxiv.org/abs/hep-th/0103093}{arXiv:hep-th/0103093}]. 

\bibitem{Yoon} Y. Yoon, ``Conformally Coupled Induced Gravity with Gradient Torsion", 	\href{https://www.sciencedirect.com/science/article/pii/S0370157301000308?via\%3Dihub}{Phys. Rev. D {\bf59}, 127501 (1999)} [\href{https://arxiv.org/abs/gr-qc/9904018}{arXiv:gr-qc/9904018}]. 
\bibitem{Obukhov} Y. N. Obukhov, ``Poincare gauge gravity: selected topics", \href{https://www.worldscientific.com/doi/abs/10.1142/S021988780600103X}{Int. J. Geom. Meth. Mod. Phys. {\bf03}, 95 (2006)} [\href{https://arxiv.org/abs/gr-qc/0601090}{arXiv:gr-qc/0601090}]. 

\bibitem{Hammad2012} F. Hammad, ``An entropy functional for Riemann-Cartan space-times", \href{https://link.springer.com/article/10.1007%2Fs10773-011-0913-9}{Int. J. Theor. Phys. {\bf51}, 362 (2012)}; \href{https://link.springer.com/article/10.1007%2Fs10773-013-1806-x}{Erratum-ibid. {\bf52}, 4592 (2013)} [\href{https://arxiv.org/abs/1202.0966}{arXiv:1202.0966v3}].

\bibitem{ChakrabortyDey} S. Chakraborty and R. Dey, ``Noether current, black hole entropy and spacetime torsion", \href{https://www.sciencedirect.com/science/article/pii/S0370269318307962?via%3Dihub}{Phys. Lett. B {\bf786}, 432 (2018)
} [\href{https://arxiv.org/abs/1806.05840}{arXiv:1806.05840}].

\bibitem{ConformalIssue} I. Quiros, R. Garcia-Salcedo, J. E. M. Aguilar, and T. Matos ``The conformal transformation's controversy: what are we missing?", \href{https://link.springer.com/article/10.1007\%2Fs10714-012-1484-7} {Gen, Rel. Gravit. {\bf45}, 489 (2013)} [\href{https://arxiv.org/abs/1108.5857}{arXiv:1108.5857}].

\bibitem{FaraoniPrainZambrano} V. Faraoni, A. Prain and A. F. Zambrano Moreno, ``Black holes and wormholes subject to conformal mappings'', \href{https://journals.aps.org/prd/abstract/10.1103/PhysRevD.93.024005}{Phys. Rev. D {\bf 93}, 024005 (2016)} [\href{https://arxiv.org/abs/1509.04129}{arXiv:1509.04129}].

\bibitem{Hammad3} F. Hammad, ``Revisiting black holes and wormholes under Weyl transformations", \href{https://journals.aps.org/prd/abstract/10.1103/PhysRevD.97.124015}{Phys. Rev. D {\bf97}, 124015 (2018)} [\href{https://arxiv.org/abs/1806.01388}{arXiv:1806.01388}].

\bibitem{Hammad1} F. Hammad, ``Conformal mapping of the Misner-Sharp mass from gravitational collapse", \href{https://www.worldscientific.com/doi/abs/10.1142/S0218271816500814}{Int. J. Mod. Phys. D {\bf25}, 1650081 (2016)} [\href{https://arxiv.org/abs/1610.02951}{arXiv:1610.02951}].

\bibitem{Hammad2} F. Hammad, ``More on the conformal mapping of quasi-local masses: The Hawking-Hayward case", \href{http://iopscience.iop.org/article/10.1088/0264-9381/33/23/235016/meta}{Class. Quantum Grav. {\bf33}, 235016 (2016)} [\href{https://arxiv.org/abs/1611.03484}{arXiv:1611.03484}].

\end{thebibliography}
\end{document}